\documentclass[5p,times,authoryear]{elsarticle}

\biboptions{square,comma}
\setcitestyle{numbers}

\usepackage{amsmath}  
\usepackage{color,graphicx}
\usepackage{txfonts}  
\usepackage{epstopdf}
\usepackage{mathptmx}                
\usepackage{dcolumn}                 
\usepackage{bm}                      
\usepackage{ulem,soul}
\usepackage{lineno,hyperref}

\hypersetup{%
    pdfsubject=Paper,
    pdfkeywords={nuclear physics} 
    unicode=true,
    breaklinks=true,
     colorlinks   = true,
     linkcolor = blue,
     citecolor = blue,
     menucolor = blue,
     urlcolor = blue
}

\graphicspath{{./figures/}}

\newcommand{\sat}{\mathrm{sat}}
\newcommand{\sym}{\mathrm{sym}}
\newcommand{\tov}{\mathrm{TOV}}
\newcommand{\qyc}{\mathrm{Qyc}}
\newcommand{\had}{\mathrm{Had}}
\newcommand{\nuc}{\mathrm{nuc}}

\newcommand{\cl}{\mathrm{CL}}
\newcommand{\fo}{\mathrm{FO}}

\newcommand{\delete}{\bgroup\markoverwith{\textcolor{red}{\rule[0.5ex]{2pt}{1pt}}}\ULon}

\begin{document}

\title{Impact of massive neutron star radii on the nature of phase transitions in dense matter}




\author{Rahul Somasundaram and J\'er\^ome Margueron}
\address{Univ Lyon, Univ Claude Bernard Lyon 1, CNRS/IN2P3, IP2I Lyon, UMR 5822, F-69622, Villeurbanne, France}

\date{\today}

\begin{abstract}
The last few years have seen tremendous progress in the observation of the global properties of neutron stars (NSs), e.g. masses, radii and tidal deformabilities. 
Such properties provide information about possible phase transitions in the inner cores of NSs, provided the connection between observed masses and radii and the equation of state (EoS) is well understood.
We focus the present study on first-order phase transitions, which often softens the EoS and consequently reduces the maximum mass as well as the radii of NSs.
Here, we challenge this conventional expectation by constructing explicit examples of EoSs undergoing a first-order phase transition, but which are much stiffer that their purely hadronic counterparts. We also provide comparisons with the recently proposed quarkyonic EoS which suggests a strong repulsion in the core of NSs, and we show that their stiffness can be realistically masqueraded by first-order phase transitions to exotic matter.   
\end{abstract}

\maketitle



\textit{Introduction}.— Neutron Stars (NSs) are one of the most fascinating objects in the universe, providing us with a wealth of nuclear and astrophysical data~\cite{Lattimer:2004}. Recent radio, x-ray and gravitational wave observations~\cite{TheLIGOScientific:2017,Miller:2019,Riley:2019,Cromartie:2019} have provided valuable new insights into the Equation of State (EoS) of dense matter. See Refs.~\cite{Bogdanov:2019b,Abbott:2018,Tews:2018} for a recent review. Additionally, there has been a renewed and significant effort to address the question of the composition of the inner core of NSs, with the plausible existence of deconfined quark matter (QM). The latest result of the NICER observation concerning the radius estimation of one of the most massive NSs presently known~\cite{NICER2021} lays down the question of the squeezability of very dense matter in the core of NSs: it suggests that the exotic phase in very dense matter is repulsive enough to equilibrate against the strong gravitational fields and maintain a relatively large radius in massive NSs.
The question of the interpretation of the massive star's radius measurement in terms of the nature of the possible phase transition in dense matter is thus very timely, and the answer is maybe not straight forward. The purpose of the present work is to address this question and explore the consequences of the onset of a first-order phase transition to an exotic state of matter. We also confront it to the predictions of the recent quarkyonic model~\cite{McLerran:2018, Jeong:2019,Sen:2020,Duarte:2020, Zhao:2020, Margueron_prep}.

The existence of QM in the cores of NSs has a long history starting from the early works by Seidov~\cite{1971SvA}, Bodmer \cite{Bodmer:1971} and Witten \cite{Witten:1984}. In particular, Ref.~\cite{Witten:1984} explored the strange quark matter hypothesis stating that the absolute ground-state of matter may be composed of u,d,s quarks instead of the observed u,d matter that build nucleons. Since then, typical questions such as the nature of the transition, its location in the space of thermodynamic variables and implications for the resulting EoS have attracted a lot of attention \cite{1983A&A...126..121S,Lindblom:1998,Zdunik:2013,Chamel:2013,Most:2018,Weih:2019}. 
Theoretical modeling of QM has been investigated from the simple MIT bag model to more advanced field-theory based NJL models, see Ref.~\cite{Buballa:2003} for a review of the latter.
Besides these "microscopic approaches", it was originally proposed by Zdunik et. al.~\cite{Zdunik:2013} (see also Ref.~\cite{Alford:2013}) to investigate a more agnostic type of modeling where the first-order phase transition is described in terms of physical quantities, instead of coupling constants.
In this framework, the EoS is
\begin{equation}
  p(n) =
  \begin{cases}
    p_\had(n) & \text{if $n < n_\fo$} \\
    p_{\fo}  & \text{if $ n_\fo < n < n_\fo + \delta n_\fo $} \\
     p_{\fo} + c^2_{\fo} \big(\rho(n) - \rho_\fo \big) & \text{if $n > n_\fo + \delta n_\fo $}
  \end{cases}
  \label{eq:first_order}
\end{equation}
where $p(n)$ is the pressure as a function of the baryon number density, with $p_\had (n)$ denoting the purely hadronic case that is valid below the transition density $n_\fo$ (the subscript refers to First-Order). Also, $p_\fo$ is the constant pressure in the mixed phase which exists in the density interval $\delta n_\fo$, with $p_\fo=p_\had(n_\fo)$. The variable $\rho(n)$ is the energy density and $\rho_\fo$ is the energy density at $n = n_\fo + \delta n_\fo$. Finally, $c_\fo$ is the sound speed in the exotic matter (EM) phase, which is assumed to be a constant, at least for the explored densities just after the FOPT. Note that in this approach there is no assumption about the nature of the EM, which is not necessarily deconfined QM, and thus this approach is quite
agnostic regarding the composition of the exotic phase.

In Eq.~\eqref{eq:first_order} the pressure remains constant as a function of $n$, during the gap $\delta n_\fo$. This leads to the so called \textit{softening} of the EoS. Although one might think that such soft EoSs may not support NSs with $M\geq 2M_{\odot}$ (a condition which is required by radio observations of heavy pulsars, see Refs.~ \cite{Fonseca:2016,Arzoumanian:2017,Linares:2018,Cromartie:2019} for the most recent of them), studies have shown that FOPTs can lead to maximum masses $M_\tov \sim 2.5 M_{\odot}$~\cite{Alford:2013}. Additionally, large radii ($\sim 14$~km) for massive NSs have been constructed in previous works, see for instance Refs.~\cite{Han:2020,Chamel:2013}. In particular, the authors of Ref.~\cite{Han:2020} have constrained the properties of FOPTs by imposing upper and lower bounds on $M_\tov$ along with bounds on the radii of massive NSs. In Ref.~\cite{Chamel:2013}, the authors obtain radii as large as $15.8$~km for maximally massive stars. However this value decreases significantly if the condition of thermodynamic stability is imposed, i.e. the condition that the exotic phase should have a lower free energy per baryon than the nucleonic phase. 
This stability suggests that the nucleonic free energy represents a viable solution for the ground-state of matter at all densities, even for the largest ones found after the phase transition. Such a condition might be over constraining since it may reject solutions which can be physical after the phase transition. Since the break-down density of the nucleonic solution is not well known, we fix it to be located just after the phase transition, for $n>n_\fo+\delta n_\fo$.

The softening of the EoS associated to a FOPT is contrasted by the quarkyonic model (QycM) recently applied to compact stars~\cite{McLerran:2018, Jeong:2019,Sen:2020,Duarte:2020, Zhao:2020, Margueron_prep}.
The multi-component momentum space shell-structure of the QycM results in a rapid increase in pressure as a function of $n$ upon the onset of quarks. This induces an hardening of the EoS in a density interval associated to the cross-over transition between hadronic and quark matter. At higher density, the sound speed decreases again, producing the expected softening of the EoS, but at densities that may not be explored in nature.
The qualitative features of QycM seems thus to be opposite to the ones of the dense matter FOPT.



\textit{Model for the nucleon phase}.—The purpose of this letter is to challenge this \textsl{a priori} difference between FOPT and QycM by testing to which extent FOPTs can also predict \textsl{hard} dense matter EoSs. Our construction of the first-order phase transition is done using Eq.~\eqref{eq:first_order}. Such a construction, below the transition point requires a model for the hadronic EoS. The hadronic EoS can, in principle, include more than just nucleons, e.g. nucleon resonances, hyperons, meson condensates. For simplicity, we consider in the following only nucleonic EoS below the phase transition.
For this purely nucleonic EoS $\rho_{\textrm{Nuc}}(p)$, we use the meta-model (MM) introduced in Refs.~\cite{Margueron:2017a, Margueron:2017b} (See also Refs.~\cite{Somasundaram:2020,Guven:2020}).
This is a density functional approach that allows one to incorporate nuclear physics knowledge encoded in the Nuclear Empirical Parameters (NEPs). These parameters are defined via a Taylor expansion of the energy per particle in symmetric matter, $e_{\sat}$ and the symmetry energy, $e_{\sym}$ about saturation density, $n_\sat$

\begin{eqnarray}
&&\hspace{-1cm}e_{\sat}(n) = E_\sat + \frac{1}{2} K_\sat x^2 + \frac{1}{3!} Q_\sat x^3 + \frac{1}{4!} Z_\sat x^4 + \dots \, , \\
\label{eq:sym}
&&\hspace{-1cm}e_{\sym}(n) = E_\sym + L_\sym x + \frac{1}{2} K_\sym x^2 \nonumber \\  
&&+  \frac{1}{3!}  Q_\sym x^3 + \frac{1}{4!} Z_\sym x^4 + \dots \, ,
\end{eqnarray}
where $x \equiv (n-n_\sat)/(3n_\sat)$ is the expansion parameter. By varying the empirical parameters within their uncertainties, the MM is able to reproduce the EoSs predicted by a large number of existing nucleonic models~\cite{Margueron:2017a,Margueron:2017b}. 

\textit{The quarkyonic model}.— After the conjecture of the QycM by McLerran and Pisarski~\cite{McLerran:2007}, a model for Qyc matter has been developed to produce an EoS relevant to NSs~\cite{McLerran:2018}.
By considering the case of pure neutron matter (PNM) rather than beta-equilibrated matter, McLerran and Reddy have shown that the resulting EoS is stiff by nature and thereby attractive to NS phenomenology.
Following their study, extensions have been proposed to incorporate beta-equilibrium by Refs.~\cite{Zhao:2020, Margueron_prep}, and here we follow the proposal of Ref.~\cite{Margueron_prep}. 

As in Ref.~\cite{McLerran:2018}, the thickness of the shell inside which the nucleons reside is given by $\Delta_{\qyc}$, which is defined as
\begin{eqnarray}
\Delta_{\qyc} = \frac{\Lambda_{\qyc}^3}{k^2_{F_N}} + \kappa_{\qyc} \frac{\Lambda_{\qyc}}{N^2_c},
\end{eqnarray}
where $k_{F_N}$ is the nucleon Fermi momentum. There are two parameters: the Qyc scale $\Lambda_{\qyc} \approx 250-300$~MeV, which is similar to the QCD scale, and $\kappa_{\qyc} \approx 0.3$. 

The energy density contributions from the nucleons and the quarks are given by
\begin{eqnarray}
\rho_N &=& 2 \sum_{i=n,p} \int_{k^{\min}_{F_i}}^{k_{F_i}} \frac{d^3k}{(2 \pi)^3} \sqrt{ k^2 + M^2_N} + V_N (k_{F_n},k_{F_p}), \nonumber \\
\rho_Q &=& 2 \sum_{q=u,d} N_c \int_{0}^{k_{F_q}} \frac{d^3k}{(2 \pi)^3} \sqrt{ k^2 + M^2_Q},
\end{eqnarray}
where $k^{\min}_{F_i} \equiv N_c k_{F_Q} (1 \pm \delta_N)^{1/3}$, where $\delta_N$ is the nucleonic asymmetry and the momentum $k_{F_Q}$ is defined as 
\begin{eqnarray}
k_{F_Q} = \frac{k_{F_N}-\Delta_{\qyc}}{N_c} \Theta (k_{F_N}-\Delta_{\qyc}),
\end{eqnarray}
The nucleonic residual interaction $V_N$ is taken from the MM approach discussed earlier. Note that, under the assumption that chiral symmetry remains broken, we take $M_Q = M_N/N_c$ even in the quarkyonic matter. 
Having thus established the EoS of quarkyonic matter for arbitrary isospin asymmetries, the condition of beta-equilibrium can now be trivially imposed \cite{Margueron_prep}.

\begin{figure}[t!]
    \centering
    \includegraphics[width=0.99\columnwidth]{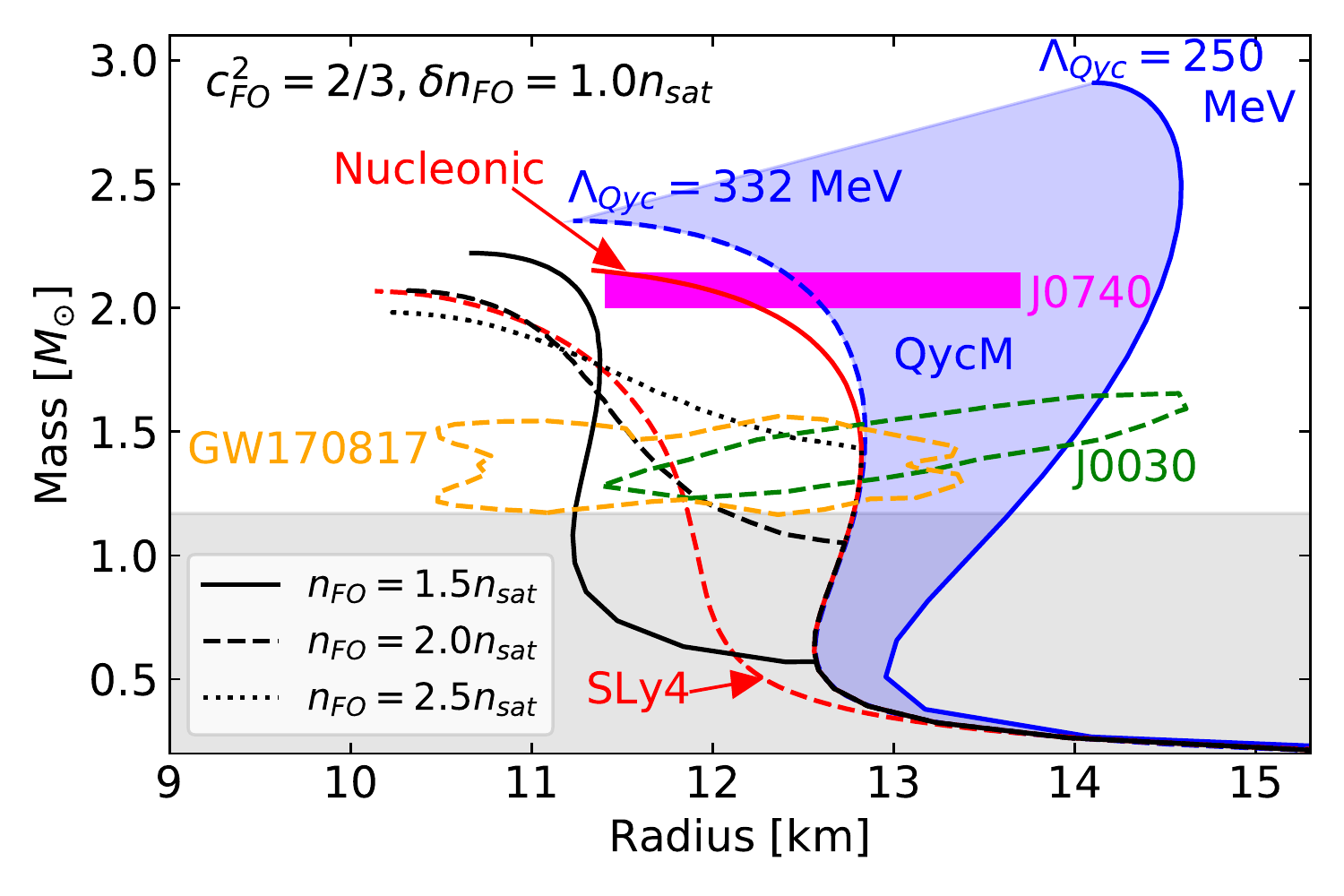}
    \includegraphics[width=0.99\columnwidth]{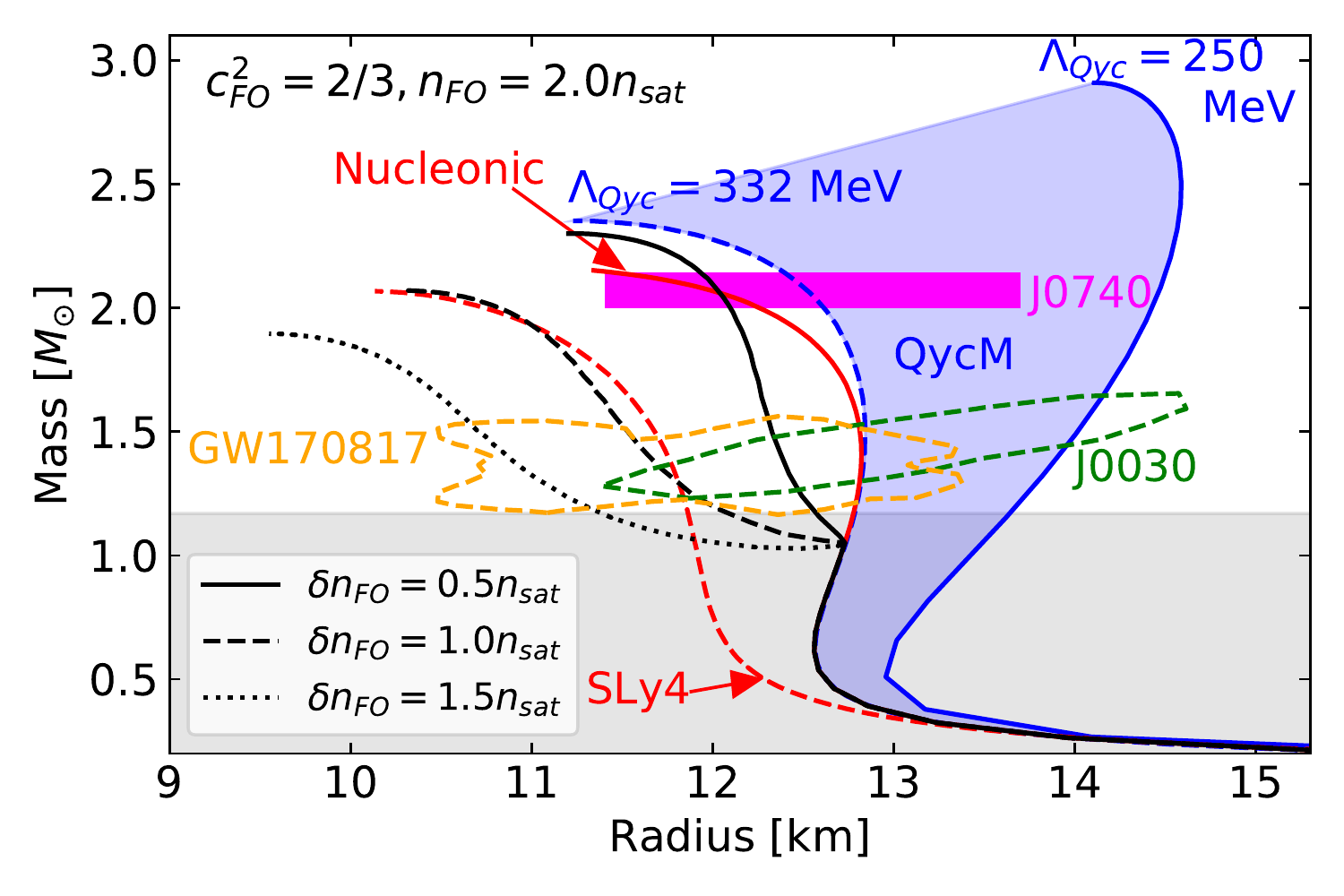}
    \includegraphics[width=0.99\columnwidth]{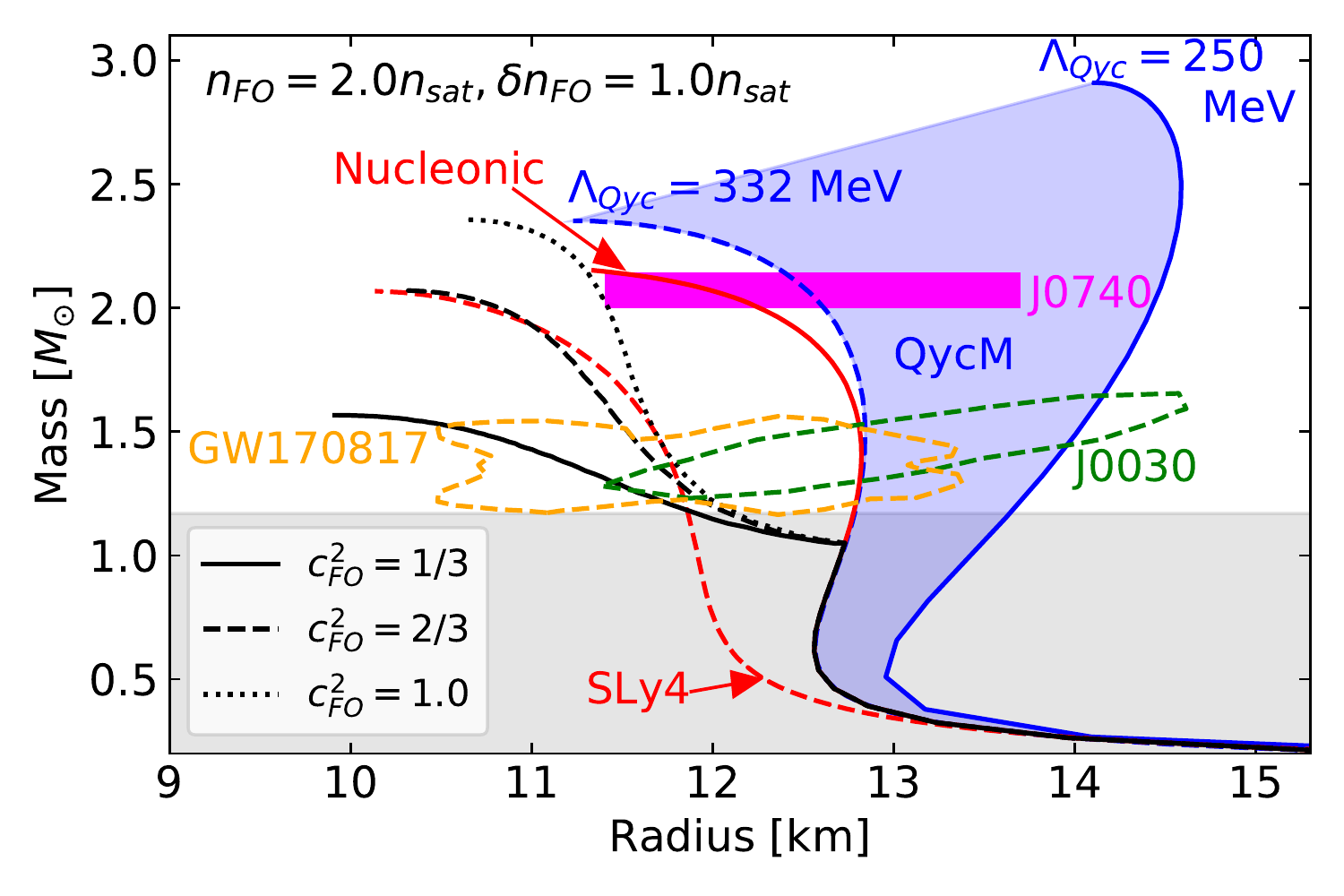}
    \caption{The mass-radius curves for various FOPT EOSs in the \textit{soft} case, shown in black with the different line-styles referring to different values of the parameters governing the FOPT. The dashed red line displays the purely nucleonic SLy4 EoS and the solid one is a more repulsive nucleonic EoS which matches with the NICER's constraints (see text).  The blue lines depict area explored by the Quarkyonic EoSs. We also show constraints from GW170817 (yellow contour), NICER's PSR J0030+0451 observation (green contour) and NICER's PSR J0740+6620 measurement (magenta band).}
    \label{fig:soft_FO}
\end{figure}

\begin{figure}[t!]
    \centering
    \includegraphics[width=0.99\columnwidth]{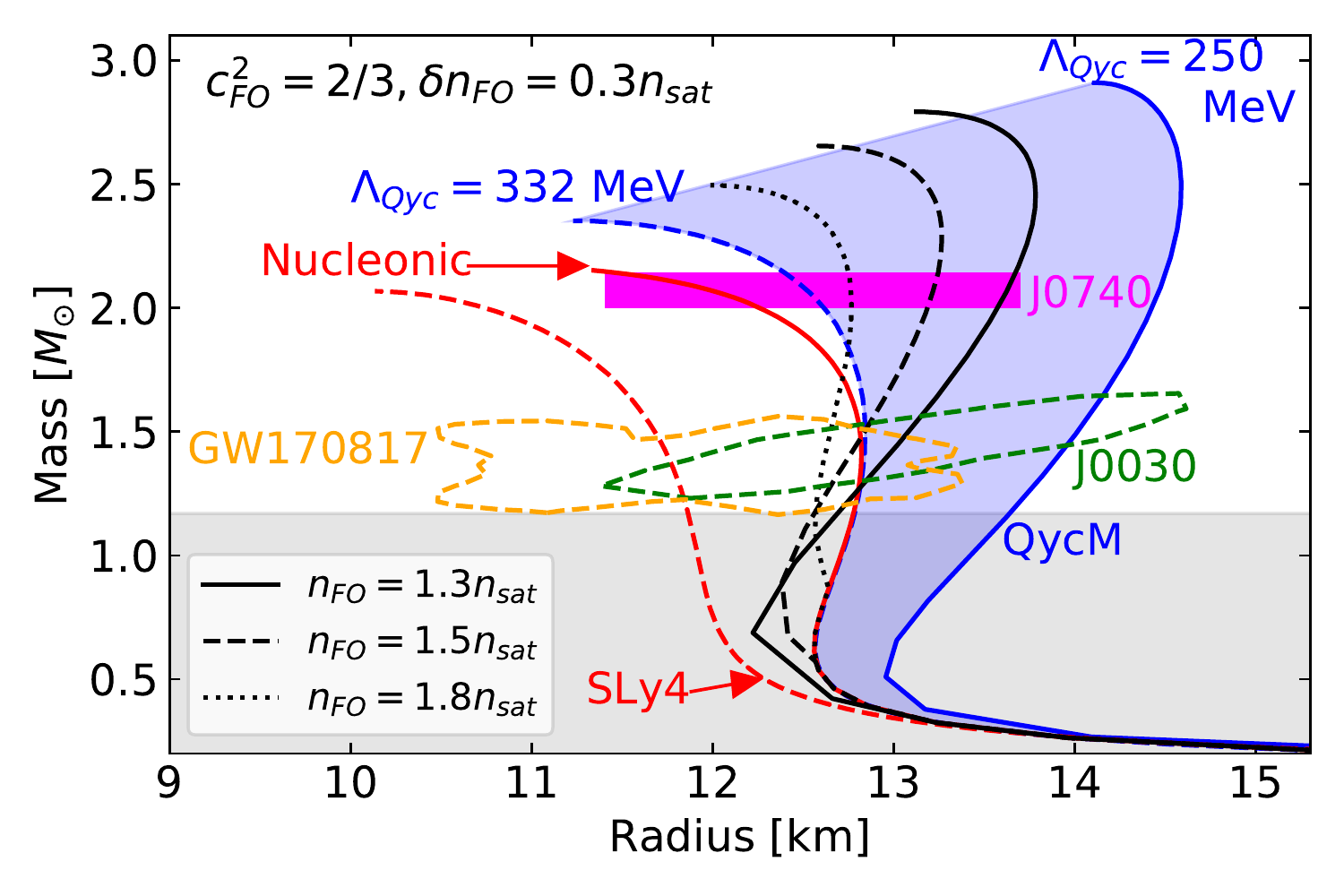}
    \includegraphics[width=0.99\columnwidth]{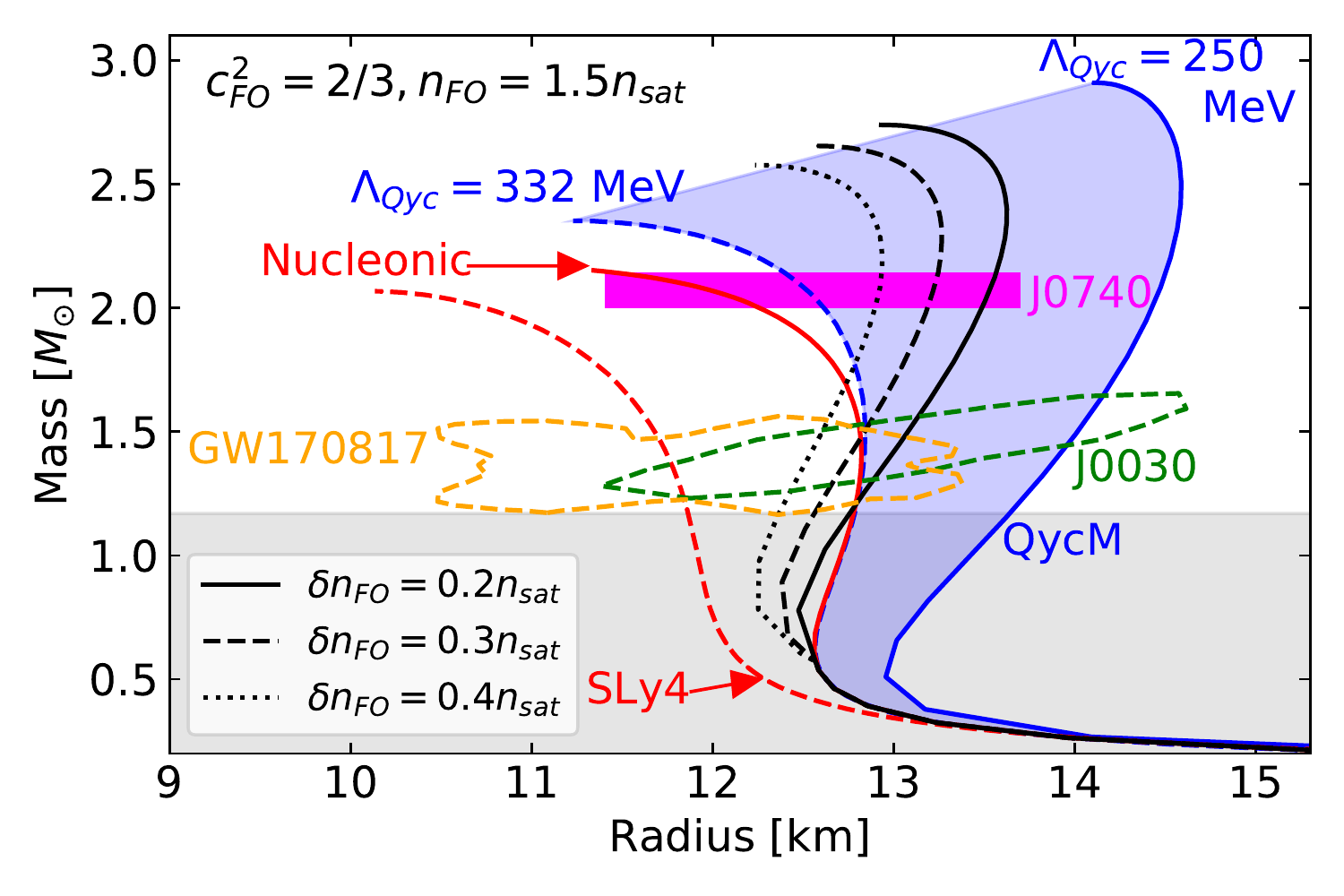}
    \includegraphics[width=0.99\columnwidth]{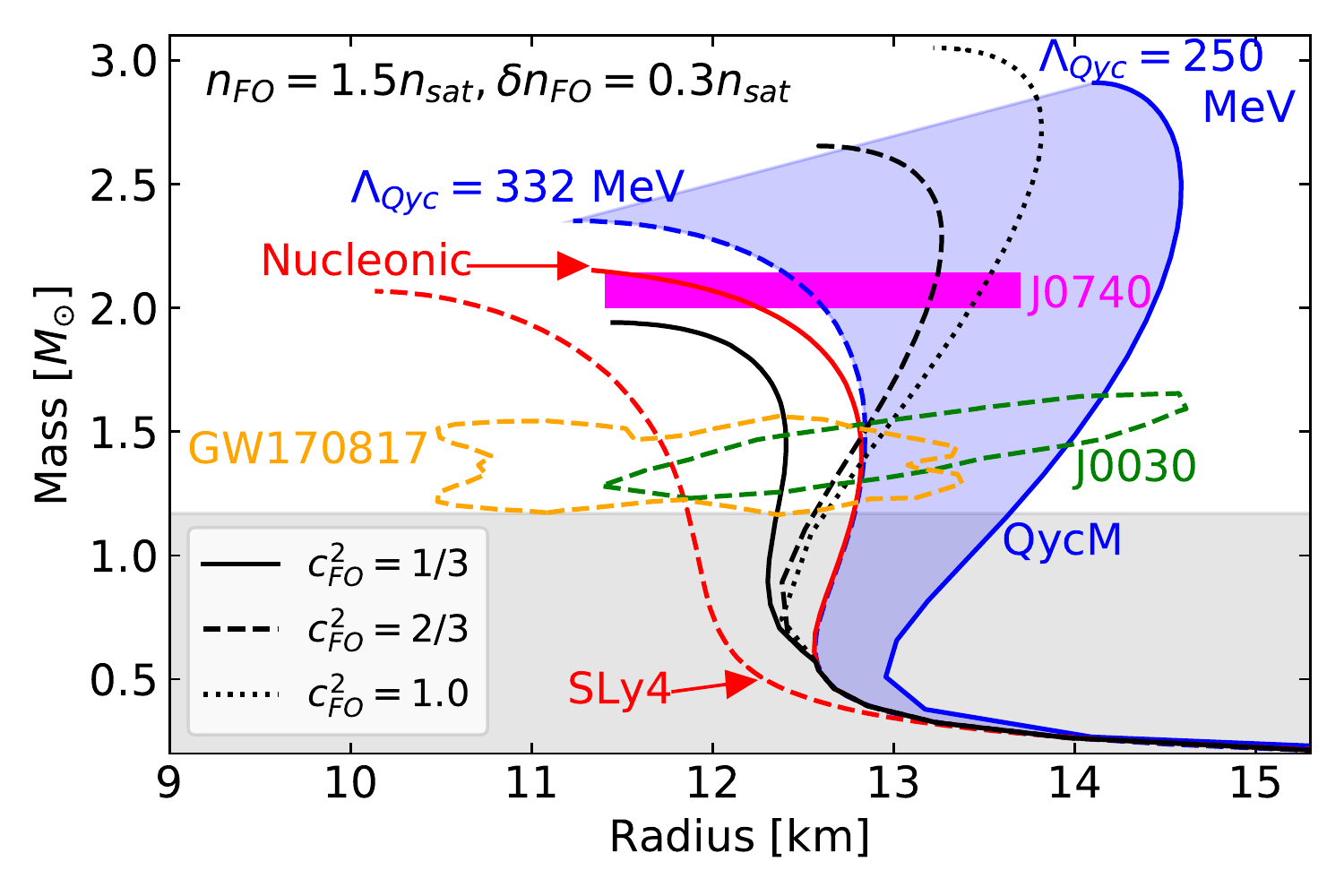}
    \caption{Same as Fig.~\ref{fig:soft_FO} for FOPT EoSs in the \textit{stiff} case.}
    \label{fig:hard_FO}
\end{figure}

\textit{Results (a)}.—In Fig.~\ref{fig:soft_FO} we show various examples of EoSs with a FOPT as described by Eq.~\eqref{eq:first_order}. These are shown as the black lines with different line-styles. We consider the purely nucleonic SLy4 EoS~\cite{Chabanat:1998} shown as the red dashed line. Interestingly, the SLy4 EoS is quite soft and it predicts radii lower than the recent NICER radius estimate~\cite{NICER2021} of PSR J0740+6620 shown in magenta (68\% CL). In order to increase the repulsion and create a stiffer nucleonic EoS, the value of $K_{\sym}$ (see Eq.\eqref{eq:sym})  was increased from $-120$~MeV to $125$~MeV and the result is shown as the solid red line which is consistent with the NICER result. In the following we shall always provide comparisons with this EoS and thus we have labelled it simply as 'Nucleonic'.  
Additionally, we show two extreme instances of quarkyonic EoSs as solid and dashed blue lines and fill up the region which would be occupied by a QycM with an choice for the parameter $\Lambda_{\qyc}$ in between the two values,
250 and $332$~MeV. These values are chosen so that they represent, in a very approximate manner, the limits of a band that encapsulates various possible quarkyonic EoSs. The value of $\Lambda_{\qyc} = 332$ MeV implies that the onset of quarks occurs at a density of $0.33$~fm$^{-3}$, whereas for $\Lambda_{\qyc} = 250$ MeV, the quarks appear at $0.14$~fm$^{-3}$. Thus, while the former corresponds to what one may consider as a typical value of the quarkyonic transition density, the latter serves only as an example of an extreme case where the quarks start appearing around the saturation density. We also show existing astrophysical data on NS masses and radii. The yellow contour depicts constraints from the GW170817 event obtained from Ref.~\cite{Abbott:2018}, the green one is obtained from the NICER observation reported in Ref.~\cite{Miller:2019} and the magenta band corresponds to the latest NICER observation of PSR J0740+6620 \cite{NICER2021}. All measurements are reported at the 68\% CL.

In Fig.~\ref{fig:soft_FO}, various values of the FOPT parameters are considered. Recall that $n_\fo$ is the baryon number density at the transition point, $\delta n_\fo$ is the transition gap and $c^2_\fo$ is the square of the sound speed in the exotic matter present in NSs after the phase transition. Here, we consider values for these parameters in the ranges: $n_\fo = (1.5 - 2.5) n_\sat$, $\delta n_\fo = (0.5 - 1.5) n_\sat$, and $c^2_\fo = 1/3 - 1$. Note that some of these EoSs can support more massive NSs than the most massive one that can be supported by the purely nucleonic EoS.
It was indeed already noticed in Ref.~\cite{Margueron_prep} that the Qyc model can change a nucleonic EoS failing to get the observed $2M_\odot$ into an EoS passing over this limit.
However, the generic feature of all these first-order EoSs is that they predict smaller 
radii compared to the nucleonic and quarkyonic EoSs. Thus the first-order EoSs considered here clearly illustrate the prevalent picture mentioned earlier, i.e. first-order phase transitions soften the EoS and lead to smaller radii~\cite{Zdunik:2013,Alford:2013,Chamel:2013}. Also note that all FOPT curves (except the solid one in the middle panel) predict smaller radii than the recent NICER estimate shown in magenta, and would therefore be simply rejected by this new observation.

In Fig.~\ref{fig:soft_FO}, the parameters controlling the transition point where chosen to 
coincide with typical values expected for a transition to QM, i.e. the transition does not occur at very low densities ($n_\fo \gtrsim 2.0 n_{sat}$) and the sound speed is close to the expected conformal limit in QM, $c_{\cl}^2=1/3$. This is what leads to the 
consensus
that FOPTs lead to soft EoSs. However, we argue that 
another set of values, with $n_\fo \lesssim 2.0 n_{sat}$, $c^2_\fo \gtrsim 0.6$ and $\delta n_\fo \approx 0.3 n_{sat}$, cannot be ignored just because they result in the appearance of quarks (or any other exotic matter) at relatively low densities. The first reasoning is that these sets of parameter values results in EoSs 
that could not be excluded by experimental or observational constraints.
Further, a recent study \cite{Xie:2020} estimating the posterior distribution over these parameters via a Bayesian analysis of astrophysical data has shown that the most probable values are $n_\fo \approx 1.6 n_{sat}$ and $c^2_\fo \approx 0.95$. This shows that parameters resulting in an early phase transition to EM with large sound speeds should not be overlooked and, additionally, they may even be favoured by astrophysical observations.


With such a motivation, in Fig.~\ref{fig:hard_FO} we construct examples of EoSs with a FOPT that are relatively stiff. We have considered low values for the transition point $n_\fo \leq 1.8 n_\sat$, small density gaps $\delta n_\fo = 0.2 - 0.4 n_\sat$ and again, a large prior for the EM sound speed $c^2_\fo = 1/3$, $2/3$, $1$. As before, we contrast these EoSs against the nucleonic EoS (in solid red) and two $\qyc$ EoSs (in blue) with $\Lambda_\qyc = 250$ and $332$~MeV. Several of the EoSs with FOPTs lie inside the blue band encapsulating various possible quarkyonic EoSs. For instance, in the middle panel ($c^2_\fo=2/3$ and $n_\fo = 1.5 n_\sat$), the first-order EoSs with $\delta n_\fo = 0.2n_\sat$ and $\delta n_\fo =0.3 n_\sat$ give radii that are comfortably larger than those corresponding to the quarkyonic EoS (with $\Lambda_\qyc = 332$~MeV) and the nucleonic EoS for $M \gtrsim 1.5 M_{\odot}$. The EoS with $\delta n_\fo =0.4 n_{\sat}$ predicts larger radii only in the range $M \gtrsim 1.75 M_{\odot}$. Similar comments can be made for the EoSs with a FOPT shown in the other panels with the lone exception being the EoS with $c^2_\fo = 1/3$ (bottom panel) that predicts lower radii for all NSs. This demonstrates that EoSs with FOPTs can give radii comparable to and larger than those resulting from quarkyonic and nucleonic EoSs at the condition that the sound speed in the EM exceeds the conformal limit in EM. We also observe that all EoSs with a FOPT (except the one with $c^2_\fo = 1/3$) are compatible with the considered astrophysical observations, most notably the NICER radius observation of the most massive NS known (magenta band).   

Let us also note that all the first-order EoSs shown in Fig.~\ref{fig:hard_FO} do lead to smaller radii at very low NS masses ($ \approx 1 M_{\odot}$). This is below the observed mass lower limit, $1.17 M_{\odot}$~\cite{Martinez:2015}, which is shown in the figure as the upper limit of the grey area.
A way to differentiate between the FOPT (for the parameters explored in Fig.~\ref{fig:hard_FO}) and Qyc matter would be to observe, if they exist, very low mass NSs ($ \approx 1 M_{\odot}$).
Another quantitative difference between FOPT and Qyc matter is observed for the predicted radii associated to a canonical mass NS. At around $1.5 M_{\odot}$, no FOPT could predict radii as large as the ones allowed by the Qyc model with typical $\Lambda_{Qyc} \approx 250$~MeV.
The main takeaway message illustrated by this figure is that EoSs with a FOPT can predict radii that are comparable to or larger than those corresponding to the nucleonic EoS as well as almost all the possible quarkyonic EoSs, in the range of observed NS masses.

In addition to the radii, Figs.~\ref{fig:soft_FO} and \ref{fig:hard_FO} can be analyzed in terms of the maximum masses $M_{\tov}$ explored by FOPTs and Qyc models.
It is known that FOPT can lead to large maximum masses, e.g. $M_{\tov} \approx 2.5 M_{\odot}$ with $c^2_\fo = 1$~\cite{Alford:2013}.
We show that under special choice of parameters, FOPT can even reach $M_{\tov} \approx 3 M_{\odot}$, see the bottom panel of Fig.~\ref{fig:hard_FO}. Since GW from binary NS mergers provide new constraints on the maximum mass~\cite{Abbott:2020khf}, FOPT and Qyc models can potentially be selected according to their prediction for $M_{\tov}$.

\begin{figure}
    \centering
    \includegraphics[width=0.99\columnwidth]{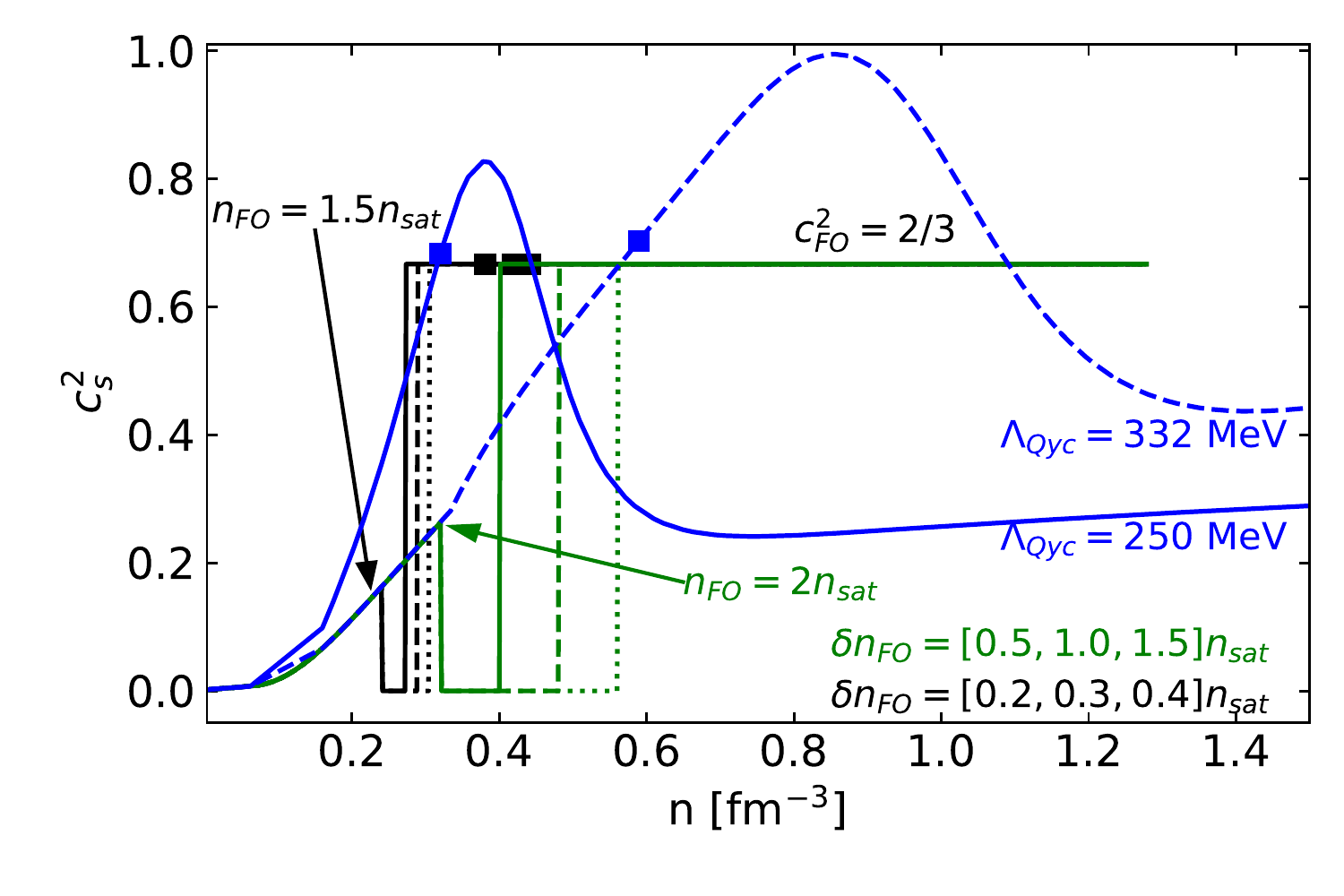}
    \caption{The sound-speed density profile for the various EoSs show in the middle panels of Figs.~\ref{fig:soft_FO} and \ref{fig:hard_FO}. The blue lines depict Quarkyonic EoSs. The black and green lines correspond to FOPTs for different values of $n_\fo$, $\delta n_\fo$ and $c^2_\fo$. The squares indicate the central density associated to a $2.1 M_{\odot}$ NS for each EoS.}
    \label{fig:sound_speed}
\end{figure}

\textit{Results (b)}.—Since FOPT and QycM can predict very similar mass-radius relation, one could ask if the sound speed predicted by these two different models share similar features as well.
In Fig.~\ref{fig:sound_speed}, the speed of sound is shown for the various EoSs considered in the middle panel of Figs.~\ref{fig:soft_FO} and \ref{fig:hard_FO}. The blue curves correspond to the $\qyc$ EoSs for the two extreme cases where $\Lambda_\qyc = 250$ and $332$~MeV, whereas the black and green ones depict the EoSs with FOPTs. The squares indicate the central density of the $2.1 M_{\odot}$ NS (the squares for the green EoSs are not shown). We see the familiar bumps in the sound speed profiles shown in blue, that characterize the quarkyonic EoS. The sound-speed density profiles associated to the FOPT models are qualitatively different from the QycM. Their density dependence is simpler since they first drop to zero for densities inside the transition domain, and they get to a constant value after the phase transition. However, the mass-radius curves generated by the FOPTs in black and the QycM in blue are quite similar, while the FOPTs in green predict different mass-radius curves (with a reduction of the radius after the phase transition).
In conclusion, Fig.~\ref{fig:hard_FO} illustrates that
the bump in the $\qyc$ EoS can be replaced with a simple, trivial structure such as a horizontal line corresponding to sound speeds around $c^2_\fo \approx 2/3$, as shown here for the FOPT in black, albeit it implies a fine tuning of the parameters.
If the value of $\delta n_\fo$ is not too large, such a replacement does not affect the mass-radius curve in a significant manner, leading to the possibility that first-order EoSs can mimic quarkyonic ones with a good accuracy. 
This remark moderates the findings of Ref.~\cite{Tan:2020}, where the authors  argued that massive NSs and stiff EoSs are likely the result of non-trivial sound speed structures.

\begin{figure}[t!]
    \centering
    \includegraphics[width=0.99\columnwidth]{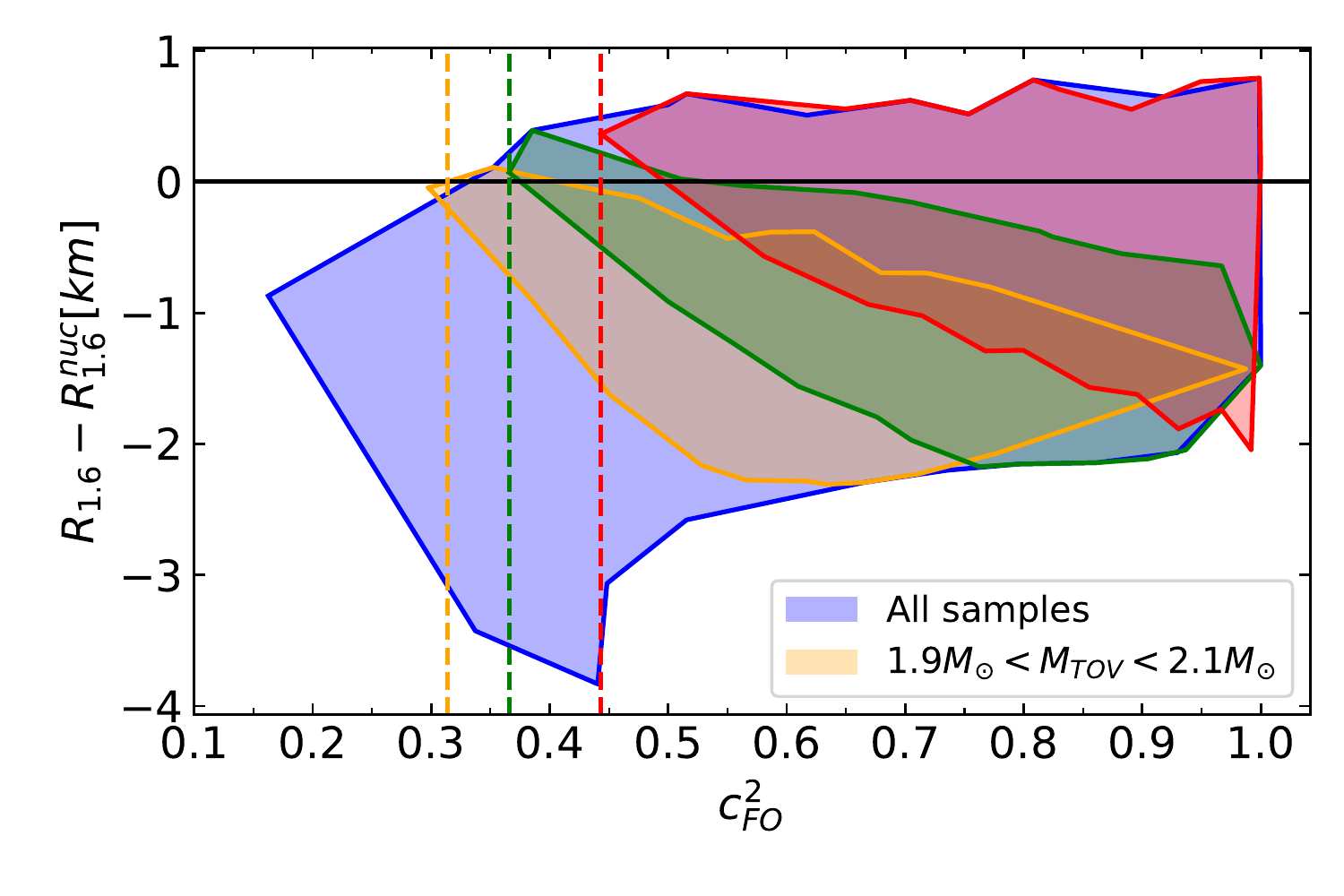}
    \includegraphics[width=0.99\columnwidth]{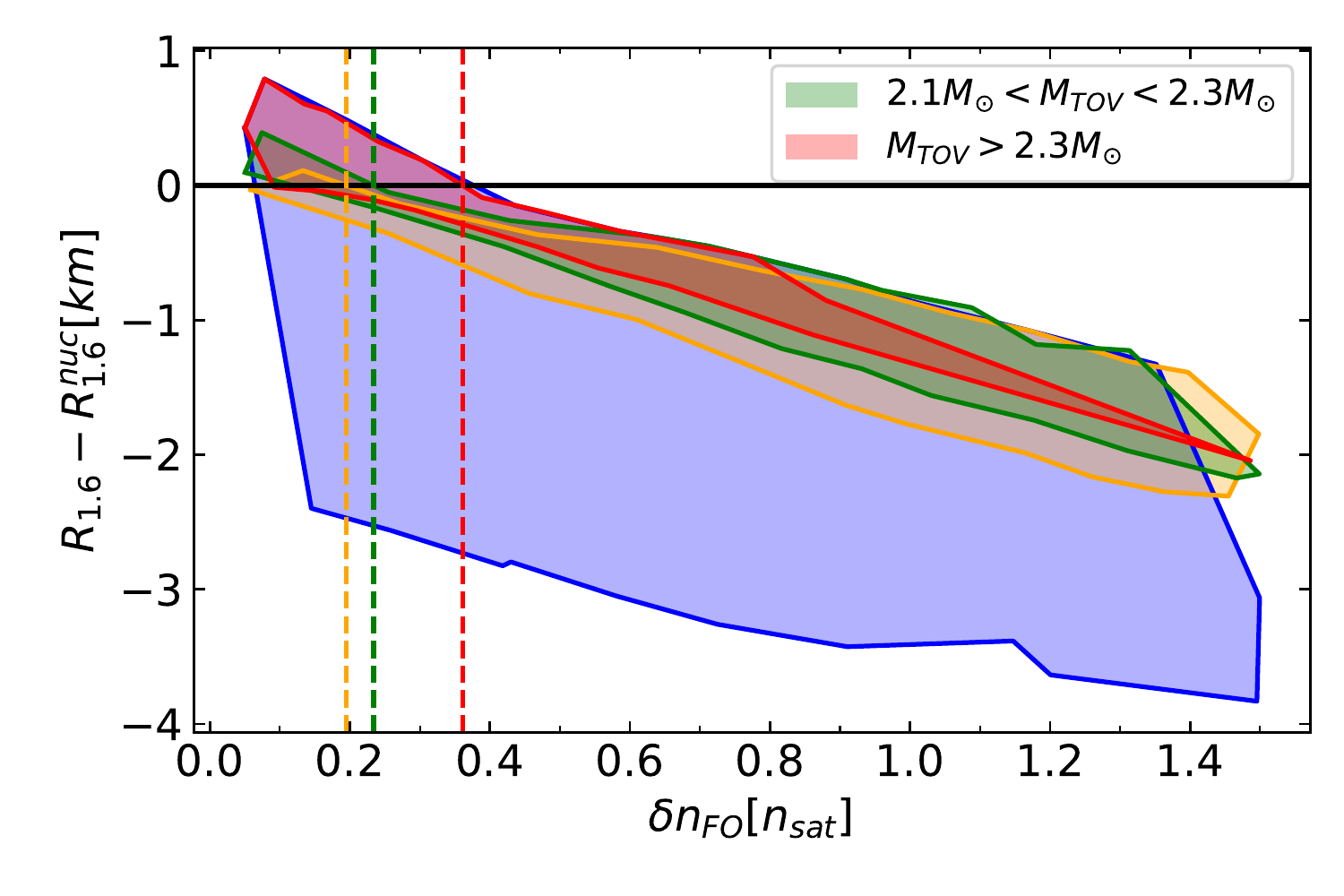}
    \includegraphics[width=0.99\columnwidth]{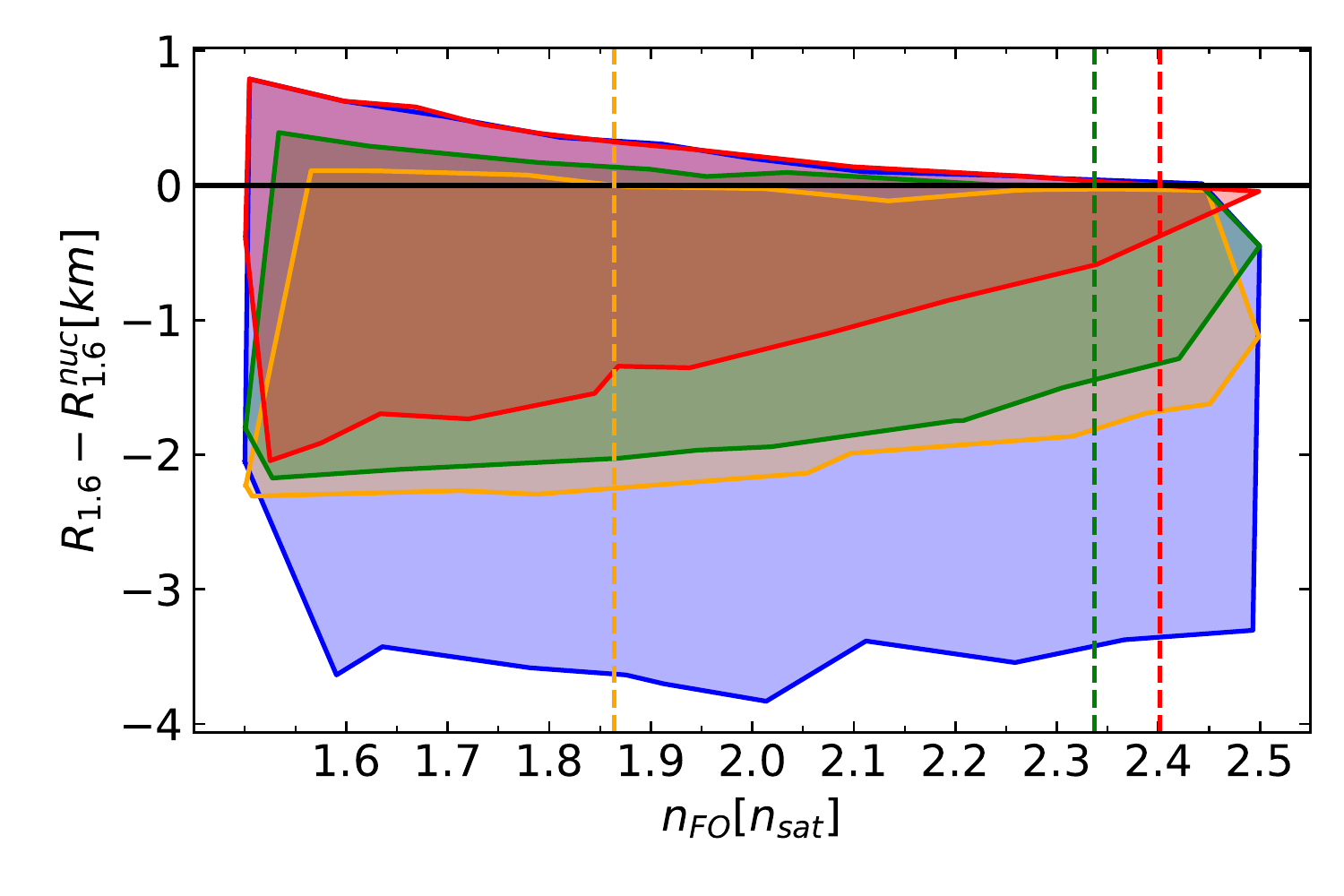}
    \caption{Plots depicting the correlation of $R_{1.6} - R^{nuc}_{1.6}$ versus the three transition parameters: $c^2_{EM}$,  $\delta n$, $n_t$. $R_{1.6}$ refers to the radius at $1.6 M_{\odot}$ resulting from a FOPT. $R^{nuc}_{1.6}$ is the same quantity but for the purely nucleonic case. The contour colors correspond to different selection rules, as indicated in the figure legend. The different vertical lines correspond to extremum values of the FOPT parameters such that the condition $R_{1.6} > R^{nuc}_{1.6}$ is satisfied. See text for more details.} 
    \label{fig:scatter_r16}
\end{figure}

\textit{Results (c)}.—Until now, we have show examples of EoSs with typical values for the FOPT parameters.
In order to understand the role played by the three parameters ($n_\fo, \delta n_\fo, c^2_\fo$) in the global properties of NS such as the radius of a $1.6 M_\odot$ NS, we now perform a more extensive analysis based on a set of 5000 EoS, where the FOPT parameters are varied in a systematical way. 
We consider a sampling of uniformly distributed parameters in the following ranges,
$c^2_\fo = [0.15,1.00]$, $\delta n_\fo = [0.05,1.50] n_{\sat}$, $n_\fo = [1.5,2.5] n_{\sat}$, and we reject all samples for which $M_\tov < 1.6 M_{\odot}$. 
For all the remaining first-order EoS samples, the quantity $R_{1.6} - R^{\nuc}_{1.6}$ is plotted as a function of the three transition parameters in Fig.~\ref{fig:scatter_r16}. Here $R_{1.6}$ refers to the radius of a $1.6 M_{\odot}$ NS resulting from a FOPT whereas $R^{\nuc}_{1.6}$ is the same quantity but for the purely nucleonic case. The results are show as contours for TOV masses given in the legend.
Notice that, by analyzing the difference $R_{1.6} - R^{\nuc}_{1.6}$, we reduce the influence of the considered nucleonic EoS on which the FOPT is built. There is however a remaining effect of the nucleonic EoS which slightly impact the numbers given below.

In the top panel of Fig.~\ref{fig:scatter_r16}, we plot $R_{1.6} - R^{\nuc}_{1.6}$ against $c^2_\fo$.
The vertical lines correspond to the minimum values of $c^2_\fo$ such that the condition $R_{1.6} > R^{\nuc}_{1.6}$ is satisfied, for different constraints on $M_\tov$ (differently colored contours).
For instance, if $M_\tov>2.1 M_\odot$ as it is likely, then $R_{1.6} > R^{\nuc}_{1.6}$ is satisfied only if $c^2_\fo \gtrsim 0.37$.  Therefore we can now confirm our earlier observation that $c^2_\fo>c^2_\cl$ allows FOPT EoS to predict larger radii than the one based on nucleonic EoS.
We also observe that the quantity $R_{1.6} - R^{\nuc}_{1.6}$ increases with the increase in $c^2_\fo$. It shows that larger sound speeds support larger TOV masses. 
For instance, if $M_\tov>2.3 M_\odot$, then $R_{1.6} > R^{\nuc}_{1.6}$ is satisfied only if $c^2_\fo \gtrsim 0.45$.

In the middle panel of Fig.~\ref{fig:scatter_r16}, we plot $R_{1.6} - R^{\nuc}_{1.6}$ against $\delta n_\fo$. The vertical lines correspond to the maximum values of $\delta n_\fo$ allowed to satisfy $R_{1.6} > R^{\nuc}_{1.6}$. Note for instance that  $R_{1.6} > R^{\nuc}_{1.6}$ (stiffer first-order EoS) is possible only if $\delta n_\fo \lesssim 0.23 n_{\sat}$ if $2.1M_\odot<M_{\tov}<2.3M_\odot$.
Unlike the top panel, EoS samples with $M_{\tov}>2.3M_\odot$ can be obtained at all values of $\delta n_\fo$.
Restricting $M_{\tov}$ to be inside small intervals but $>1.9M_\odot$ induces a tight correlation between $R_{1.6} - R^{\nuc}_{1.6}$ and $\delta n_\fo$. The correlation is even tighter for $\delta n_\fo>n_\sat$ if $M_{\tov}>2.3M_\odot$, and for $\delta n_\fo<0.6n_\sat$ if $1.9M_\odot<M_{\tov}<2.3M_\odot$.

Finally, in the bottom panel of Fig.~\ref{fig:scatter_r16}, we see that $R_{1.6} > R^{\nuc}_{1.6}$ is possible at the condition that $n_\fo \lesssim 2.3 n_{\sat}$ if $2.1M_\odot<M_{\tov}<2.3M_\odot$, and at $n_\fo \lesssim 2.4 n_{\sat}$ 
if $M_{\tov}>2.3M_\odot$. Note that these values for $n_\fo$ are above $2n_\sat$ (not so low).
Larger values for $R_{1.6}$, i.e. increasingly repulsive EoS, are obtained for the lower values for $n_\fo$, predicting as well larger TOV masses.

\textit{Conclusions}.— The purpose of this letter is to challenge the \textsl{a priori} difference between FOPT and QycM, and to show that FOPT solutions could realistically masquerade QycM and predict \textsl{hard} dense matter EoSs.
We construct explicit examples of EoSs undergoing a FOPT that are significantly stiffer than their purely hadronic counterparts, where by \textit{stiff} we typically refer to the radii of NSs. A stiffer EoS also predicts a larger TOV mass. Additionally, we show that the stiffness of such EoSs can be such that their mass-radius relations can realistically mimic the ones corresponding to the QycM. This can be seen as an extension of the results of Ref.~\cite{Alford:2004} where it was shown that the mass-radius relations of first-order EoSs are quite similar to those predicted for NSs made of purely nucleonic matter. 
We have also confirmed that large TOV masses are possible with such FOPT EoSs. These EoSs arise from a certain set of values for the FOPT parameters, and we have performed a detailed analysis of the correlation between the stiffness of the EoS and the FOPT parameters. We have argued that such FOPT parameter sets are not ruled out by present astrophysical data and may even be favoured by them, as shown in Ref.~\cite{Xie:2020}. 
These results have clear and important implications for phenomenology-based studies of the hadron-quark phase transitions in NSs and for the confrontation of EoS to the latest results from the NICER observation~\cite{NICER2021}. An interesting extension of this work would be to confront systematically the various EoSs presented in this letter to the wealth of available astrophysical data (radio, x-rays, and GW). Work along these lines is presently in progress.

R.S. is supported by the PHAST doctoral school (ED52) of \textsl{Universit\'e de Lyon}. R.S. and J.M. are both supported by the CNRS/IN2P3 NewMAC project, and are also grateful to PHAROS COST Action MP16214 and to the LABEX Lyon Institute of Origins (ANR-10-LABX-0066) of the \textsl{Universit\'e de Lyon} for its financial support within the program \textsl{Investissements d'Avenir} (ANR-11-IDEX-0007) of the French government operated by the National Research Agency (ANR).

\bibliography{biblio}

\end{document}